\documentstyle{mn}
\newif\ifAMStwofonts
\ifoldfss
  \ifCUPmtlplainloaded \else
    \NewTextAlphabet{textbfit} {cmbxti10} {}
    \NewTextAlphabet{textbfss} {cmssbx10} {}
    \NewMathAlphabet{mathbfit} {cmbxti10} {} % for math mode
    \NewMathAlphabet{mathbfss} {cmssbx10} {} %  "   "    "
  \fi
  \ifAMStwofonts
    \ifCUPmtlplainloaded \else
      \NewSymbolFont{upmath} {eurm10}
      \NewSymbolFont{AMSa} {msam10}
      \NewMathSymbol{\upi}     {0}{upmath}{19}
      \NewMathSymbol{\umu}     {0}{upmath}{16}
      \NewMathSymbol{\upartial}{0}{upmath}{40}
      \NewMathSymbol{\leqslant}{3}{AMSa}{36}
      \NewMathSymbol{\geqslant}{3}{AMSa}{3E}

    \fi
  \fi
\fi % End of OFSS

\ifnfssone
  \newmathalphabet{\mathit}
  \addtoversion{normal}{\mathit}{cmr}{m}{it}
  \addtoversion{bold}{\mathit}{cmr}{bx}{it}
  \newmathalphabet{\mathbfit} % math mode version of \textbfit{..}
  \addtoversion{normal}{\mathbfit}{cmr}{bx}{it}
  \addtoversion{bold}{\mathbfit}{cmr}{bx}{it}
  \newmathalphabet{\mathbfss} % math mode version of \textbfss{..}
  \addtoversion{normal}{\mathbfss}{cmss}{bx}{n}
  \addtoversion{bold}{\mathbfss}{cmss}{bx}{n}
  \ifAMStwofonts
    \ifCUPmtlplainloaded \else
      \UseAMStwoboldmath
      \makeatletter
      \new@mathgroup\upmath@group
      \define@mathgroup\mv@normal\upmath@group{eur}{m}{n}
      \define@mathgroup\mv@bold\upmath@group{eur}{b}{n}
      \edef\UPM{\hexnumber\upmath@group}
      \new@mathgroup\amsa@group
      \define@mathgroup\mv@normal\amsa@group{msa}{m}{n}
      \define@mathgroup\mv@bold\amsa@group{msa}{m}{n}
      \edef\AMSa{\hexnumber\amsa@group}
      \makeatother
      \mathchardef\upi="0\UPM19
      \mathchardef\umu="0\UPM16
      \mathchardef\upartial="0\UPM40
      \mathchardef\leqslant="3\AMSa36
      \mathchardef\geqslant="3\AMSa3E
    \fi
  \fi
\fi % End of NFSS release 1

\ifnfsstwo
  \DeclareMathAlphabet{\mathbfit}{OT1}{cmr}{bx}{it}
  \SetMathAlphabet\mathbfit{bold}{OT1}{cmr}{bx}{it}
  \DeclareMathAlphabet{\mathbfss}{OT1}{cmss}{bx}{n}
  \SetMathAlphabet\mathbfss{bold}{OT1}{cmss}{bx}{n}
  \ifAMStwofonts
    \ifCUPmtlplainloaded \else
      \DeclareSymbolFont{UPM}{U}{eur}{m}{n}
      \SetSymbolFont{UPM}{bold}{U}{eur}{b}{n}
      \DeclareSymbolFont{AMSa}{U}{msa}{m}{n}
      \DeclareMathSymbol{\upi}{0}{UPM}{"19}
      \DeclareMathSymbol{\umu}{0}{UPM}{"16}
      \DeclareMathSymbol{\upartial}{0}{UPM}{"40}
      \DeclareMathSymbol{\leqslant}{3}{AMSa}{"36}
      \DeclareMathSymbol{\geqslant}{3}{AMSa}{"3E}
    \fi
  \fi
\fi % End of NFSS release 2

\ifCUPmtlplainloaded \else
  \ifAMStwofonts \else % If no AMS fonts
    \def\upi{\pi}
    \def\umu{\mu}
    \def\upartial{\partial}
  \fi
\fi

\title[Variables in M5. Application of the Image Subtraction Method]
  {Variable Stars in the Globular Cluster M5. \\
  Application of the Image Subtraction Method}
\author[A. Olech et al.]
  {A.~Olech,$^{1}$
  P.R.~Wo\'zniak,$^2$ C.~Alard,$^3$ J.~Kaluzny,$^{1,4}$ and 
  I.B.~Thompson$^5$\\
  $^1$Warsaw University Observatory, Al. Ujazdowskie 4,
00-478 Warsaw, Poland (olech,jka@sirius.astrouw.edu.pl)\\
  $^2$Princeton University Observatory, Peyton Hall, NJ, 
USA (wozniak@astro.princeton.edu)\\
  $^3$DASGAL, 61 Avenue de l'Observatoire, F-75014 Paris, France\\
  $^4$Copernicus Astronomical Center,
ul. Bartycka 18, 00-716 Warsaw, Poland\\
  $^5$Carnegie Institution of Washington, 813 Santa Barbara Street,
Pasadena,CA 91101, USA (ian@ociw.edu)
}
\date{Accepted 1999 .............. Received 1999 .............}
\pagerange{\pageref{firstpage}--\pageref{lastpage}}
\pubyear{1999}

\def\LaTeX{L\kern-.36em\raise.3ex\hbox{a}\kern-.15em
    T\kern-.1667em\lower.7ex\hbox{E}\kern-.125emX}

\begin{document}

\label{firstpage}

\maketitle

\begin{abstract} 
We present $V$-band light curves of 61 variables from the core of the
globular cluster M5 obtained using a newly developed image
subtraction method (ISM). Four of these variables were previously
unknown. Only 26 variables were found in the same field using
photometry obtained with DoPHOT software.  Fourier parameters of the
ISM light curves have relative
errors up to 20 times smaller than parameters measured from DoPHOT
photometry. We conclude that the new method is very promising for
searching for variable stars in the cores of the globular clusters and
gives very accurate relative photometry with quality comparable to 
photometry obtained by HST.
We also show that the variable V104 is not an eclipsing
star as has been suggested, but is an RRc star showing non-radial
pulsations.
\end{abstract}


\section{Introduction}

The globular cluster M5 (NGC 5904) contains one of the richest sets of
variable stars in the Galaxy. It is located only two degrees
from the celestial equator and therefore it is often a target of
variable stars searches from both hemispheres. Recent
studies include those by Reid (1996), Sandquist et al. (1996),
Drissen and Shara (1998), Kaluzny et al. (1999a, 1999b) and Caputo et
al. (1999).
 
Crowding effects make it very difficult to obtain reliable photometry
of variable stars located near the central regions of globular
clusters. One solution is to use the Hubble Space Telescope (HST).
Drissen and Shara (1998) observed a $70"\times70"$ field located in the
center of M5, detecting 29 variables during a 12 hr run during which
they obtained 22 exposures. The obvious disadvantage of using the HST
for studies of variable stars in the cores of clusters is the difficulty
of obtaining the long observational runs which are essential for good
coverage of the light curves.
 
An alternative method is to use image subtraction which effectively
deals with  the problem of crowding. Recently, Alard and Lupton (1998)
presented a new method of image subtraction (ISM), which actually works
best in crowded fields since in this case all pixels in the image are
used for the determination of the convolution  kernel. Alard (1999b)
modified the code of Alard and Lupton (1998) to optimally process
regions of any stellar density.  Alard (1999a) applied this revised formalism
to the OGLE observations of Baade's Window (Wo\'zniak and Szyma\'nski 1998) and
obtained much better light curves of the microlensing events than those
measured using traditional methods such as DoPHOT software (Schechter
et al., 1993).
 
We decided  to check the usefulness of this method for observing the
light curves of variable stars in the cores of globular clusters by
analyzing  $V$-band CCD photometry of M5 taken in May and June of 1997
using the 1-m Swope telescope at Las Campanas Observatory. The main
goal of these observations was to search for main sequence eclipsing
binary stars and therefore the exposure times were long. As a result,
these data are not favorable for observations of RR Lyr variables 
since these stars are bright
enough near maximum light to be saturated on some of our CCD images.
Exposure times ranged from 300 to 500 arcsec with median seeing of 1.5
arcsec.  In spite of this, we obtained reliable photometry of 65 RR Lyr
variables  using DoPHOT (Kaluzny et al. 1999b). These 65 variables were
detected over the whole field of view of the 2048$\times$2048 CCD
camera. For the purpose of this test application of image subtraction
we have narrowed the field of view to 601$\times$601 pixels covering
$4.4'\times4.4'$ of the central part of M5. This field contains 26
variables which were detected by Kaluzny et al. (1999b). Our data set
is somewhat smaller than used by Kaluzny et al. (1999b), only
observations taken in May and June 1997 analyzed for a total of 161
frames.

\section{Data Reduction}

Initial reductions of the CCD frames consists of bias and flat field
corrections followed by the removal of cosmic ray events. This was done
with the IRAF package. The next important step is the
registration of all frames onto a common pixel grid. This was
accomplished by using the centroids of approximately 640 bright
unsaturated stars found in both the test image and the reference image.
Strong stellar density gradients in a globular clusters can result in a
fit of the coordinate transformation which is strongly dominated by
distortions in the densest part of the image.  An approximate
equalization of the number of stars used in the fit by taking the 40
brightest stars in squares of 150$\times$150 pix is sufficient to avoid
this problem. In the end we fitted the coordinate transformation with
second order polynomials.  These fits are used with a bicubic spline
interpolator to resample all of the images onto the pixel grid of the
reference image.

The preparation of the reference image warrants a few words of
discussion. The benefits of carefully constructing this image, which
will be subtracted from all of the test images, are substantial. The
average of the 20 best seeing frames with low background (after
resampling to the same pixel grid) is practically noiseless compared to
a single exposure.  This is essential if we are to approach the best
possible accuracy. For many data sets the difference frames will be
limited only by the photon noise of a single test frame.

Registered frames are then processed with the image subtraction code
described by Alard and Lupton. The reference image is degraded to the
seeing of each frame and the deviation between the two images is
minimized. Areas covered by brightest stars and variables are not
included in the fit. For a thorough explanation of the method we refer
to the original paper of Alard and Lupton (1998). Small PSF gradients
are taken into account by subdividing each frame into 3$\times$3
subframes before fitting the convolution kernel.
 
Variables are detected using the ``variability image'', an average of
the absolute values of all difference images, which contains the
accumulated contributions from all (positive and negative) variations
with respect to the reference image. The PSF shape of the stellar images
is preserved in the variability image, and therefore practically any
software for the detection of stars can provide a list of candidate
variables. Our star finding program is based on the properties of the
cross-correlation image with the approximate Gaussian model of the PSF.
The cross-correlation function, calculated here as the convolution with
the lowered Gaussian filter, has maxima at the positions of stellar
objects. Comparison of the signal with the estimated noise for each
candidate constitutes the final selection criterion. We experimented
with various sigma cuts and found a sharp transition between a regime
where new detections are still almost entirely variables and a regime
where the candidate list grows by accumulation of noise features and CCD
defects. We imposed a cutoff of 4 sigma for the detection of a
candidate variable.
 
The actual profile photometry on individual difference images is done
using the PSF for the reference image convolved with the best fit PSF
matching kernel for each test image. We modeled the first order
spatial variation of the PSF in the reference image using the code
written for the DENIS survey (Alard 1999, in preparation).

\begin{table*}
 \centering
 \begin{minipage}{240mm}
  \caption{Elements of the RR Lyrae variables in the core of M5}
  \begin{tabular}{||lrrc||lrrc||lrrc||}
\hline
\hline
Star & X[$"$] & Y[$"$] & Period~~~ & Star & X[$"$] & Y[$"$] & Period~~~ & Star & X[$"$] & Y[$"$] & Period~~~ \\
\hline
\hline
V4 & -12.3 & +73.8 & 0.450999~~~~           & V80 & -48.6 & +111.6 & 0.336691~~~~       & V114 & +29.7 & -2.6 & 0.604015~~~~\\
V5 & -7.8 & +51.6 & 0.546416~~~~          & V81 & -72.2 & -121.7 & 0.556798~~~~       & V115 & +46.2 & +4.9 & 0.614249~~~~\\
V6 & +27.2 & -46.6 & 0.547204~~~~           & V82 & -67.8 & +12.4 & 0.558208~~~~        & V116 & +46.0 & -2.3 & 0.347421~~~~\\
V13 & +11.0 & -65.4 & 0.512984~~~~  & V87 & +122.0 & -1.8 & 0.742259~~~~        & V117 & +24.9 & +2.8 & 0.335578~~~~\\
V14 & -145.6 & +103.7 & 0.486850~~~~        & V88 & +65.2 & +61.8 & 0.327669~~~~        & V118 & +20.0 & +7.6 & 0.580965~~~~\\
V16 & +91.0 & +83.9 & 0.647695~~~~  & V89 & +60.0 & +64.7 & 0.559149~~~~        & V119 & +12.3 & +15.6 & 0.551128~~~~\\
V24 & -46.8 & -71.7 & 0.478513~~~~  & V91 & -36.0 & +35.0 & 0.585423~~~~        & V123 & -0.7 & +28.6 & 0.601801~~~~\\
V26 & +21.8 & +101.5 & 0.623978~~~~ & V93 & +44.0 & -35.7 & 0.552875~~~~        & V127 & -37.1 & +9.2 & 0.544965~~~~\\
V27 & -6.7 & -59.2 & 0.471333~~~~           & V94 & -23.5 & +17.4 & 0.533101~~~~        & V128 & -36.5 & -8.9 & 0.305704~~~~\\
V33 & -21.1 & +127.5 & 0.501980~~~~ & V95 & -47.2 & +102.8 & 0.290975~~~~       & V129 & -84.2 & -88.1 & 0.553881~~~~\\
V34 & +84.3 & +59.5 & 0.567632~~~~  & V96 & -12.4 & +32.9 & 0.511861~~~~        & V130 & +77.8 & +57.4 & 0.327450~~~~\\
V35 & -12.2 & -114.7 & 0.308402~~~~ & V97 & +48.9 & -92.5 & 0.545176~~~~        & V131 & +78.1 & +54.9 & 0.281489~~~~\\
V38 & -44.2 & +117.2 & 0.470430~~~~ & V98 & +37.8 & +20.0 & 0.306085~~~~        & V133 & +107.3 & +43.7 & 0.294905~~~~\\
V44 & -102.5 & +31.1 & 0.329585~~~~ & V99 & +34.4 & -0.1 & 0.321359~~~~ & V135 & -7.5 & -50.9 & 0.627740~~~~\\
V45 & -116.7 & +65.7 & 0.617091~~~~ & V100 & +2.8 & +48.7 & 0.294395~~~~        & V137 & +45.1 & +40.1 & 0.615718~~~~\\
V47 & -75.3 & +58.1 & 0.540381~~~~  & V103 & +20.5 & -8.8 & 0.567138~~~~        & V139 & -18.2 & +31.9 & 0.300501~~~~\\
V52 & +107.9 & +35.3 & 0.501906~~~~ & V104 & -10.2 & +42.1 & 0.310930~~~~       & V160 & -49.6 & -42.4 & 0.089767~~~~\\
V54 & +30.3 & +57.2 & 0.453850~~~~  & V109 & +19.2 & +1.5 & 0.472582~~~~        & V161 & +4.9 & +51.6 & 0.331570~~~~\\
V56 & -68.9 & +96.5 & 0.533973~~~~  & V110 & +23.8 & +16.4 & 0.598528~~~~       & V162 & -44.6 & +15.3 & 0.557468~~~~\\
V57 & -30.6 & +99.7 & 0.284739~~~~  & V112 & +28.7 & -31.4 & 0.537907~~~~       & V163 & -26.2 & +44.1 & 0.600853~~~~\\
V60 & -109.7 & +8.2 & 0.285274~~~~  &      &       &       &                &      &       &       &         \\
\hline
\hline
\end{tabular}
\end{minipage}
\end{table*}

\begin{table}
%% \centering
%% \begin{minipage}{240mm}
  \caption{Fourier Elements of the three RR Lyrae variables in the core
of M5. The comparison of the relative errors of the light curves
obtained by ISM and DoPHOT.}
  \begin{tabular}{lcccccc}
\hline
\hline
Star & ${\Delta A_1}\over A_1$ & ${\Delta A_2}\over A_2$ & ${\Delta
A_3}\over A_3$ & ${\Delta A_4}\over A_4$ & $\Delta$ \\
\hline
\hline
V27 ISM & 0.006 & 0.009 & 0.011 & 0.014 & 0.031 \\
V27 DoPHOT & 0.018 & 0.036 & 0.053 & 0.082 & 0.050 \\
\hline
V54 ISM & 0.013 & 0.024 & 0.032 & 0.046 & 0.023 \\
V54 DoPHOT & 0.007 & 0.015 & 0.021 & 0.033 & 0.043 \\
\hline
V91 ISM & 0.009 & 0.017 & 0.026 & 0.045 & 0.036 \\
V91 DoPHOT & 0.086 & 0.152 & 0.277 & 0.243 & 0.155 \\
\hline
\hline
\end{tabular}
%\end{minipage}
\end{table}

\section{Results}
 
The overall quality of our data set was not very good, and the photon
noise limit was not achieved. This is due to the combined effects of a
slight nonlinearity at the level of 4\% and residual differential
refraction which just starts being noticeable in the standard V
photometric band (see Alcock et al. 1999 for a description of the
phenomenon and correction). As a result the final accuracy was about 3
times  the photon noise. Despite these imperfections, the final light
curves are very good and the power of the method to detect variables in
the core of this cluster is impressive. In addition, we expect that
many variables were lost because of heavy saturation of bright stars
near the center of the cluster.
 
In the $4.4'\times4.4'$ field covering the core of M5 we detected 61
variables. The light curves of these stars are plotted in Fig. 1, and
the coordinates and periods are
presented in Table 1. The coordinates of the previously known variables
are taken from Sawyer Hogg (1973) and Sandquist et al. (1996). The
numbering scheme for variables from V1 to V103 was taken from Sawyer
Hogg (1973) and from V104 to V159 from Caputo et al. (1999), who
rationalized many lists of variables onto a common numbering scheme.
Four of our variables are new discoveries, and are labeled V160 through
V163.  A total of 60 of our variables are RR Lyr stars, with 19 variables
belonging to Bailey type c and 41 to Bailey type ab. We found one
SX Phe variable. We present finder charts for the newly
discovered variables in Fig. 2.
 
\subsection{Comparison with the results of Kaluzny et al. (1999b)}

It is worthwhile to compare photometry measured from  difference images
presented here with values measured with the DoPHOT software. This
comparison is shown in Fig. 3, where we have plotted the light curves
of three typical variables (V27, V54 and V91) detected both by Kaluzny
et al. (1999b) and in this work. We show the DoPHOT photometry and the
ISM photometry in the first and the second panel respectively. One can
clearly see that the measurements on the right side of Fig. 3 are much
more accurate than the ones on the left side. For an objective
comparison, we fitted all of these curves with a Fourier sine series in
the form:
 
\begin{equation}
{\rm brightness} = A_0 + \sum^{8}_{j=1} A_j\cdot\sin(j\omega t + \phi_j)
\end{equation}
 
\noindent where $\omega=2\pi/P$ and $P$ is the pulsation period of the
star and calculated the relative errors of $A_j$. These
quantities are presented in Table 2. It is clear that the relative
errors are much smaller using the ISM. The result is most striking for
V91 where the relative errors of IMS are $\sim20$ times smaller than the
errors produced by DoPHOT. The smallest differences of the relative
errors between these two methods are noted for V54 but still the light
curve of this star obtained by IMS has considerably smaller scatter than
the light curve obtained with DoPHOT. The light curve of V27 shows
behavior common among RR Lyr variables detected in this work. It has
larger scatter of the observational points around maxima than around
minima. There are two factors contributing to the effect. First, many RR
Lyr found in this search are saturated near maxima, at least in some
frames. Saturated pixels are rejected in the PSF fit and convolution through
renormalization, nevertheless convolution is nonlocal and the spreading of
defects cannot be eliminated completely. The saturated pixels are
simply rejected in the DoPHOT reductions. The second factor comes from
imperfect phasing of the light curves, which manifests itself more strongly
near maxima, where light variations are steep.  

Additionally knowing the minimal and maximal magnitudes of these
variables we transformed the light curves obtained by ISM into the
relative magnitudes and then again computed the Fourier fit in form
presented in (1). Next we computed the deviation parameter defined as:
 
\begin{equation}
\Delta={1\over{N}}\cdot\sum^N_{i=1} |{\rm mag(jd_i)}-{\rm brightness(jd_i)}|
\end{equation}
 
\noindent where $N$ is the number of observations, mag(${\rm jd}_i$) is
the relative magnitude at given HJD and brightness(jd$_i$) is the
magnitude of the star for the same HJD computed from (1).
 
The $\Delta$ parameter is given for each star in the last column of
Table 2. It is clear that the ISM photometry of variables V27 and V54 is
about two times better than the DoPHOT photometry, and the improvement for variable V91 is as large as a factor of 4.3.

\subsection{Variable V104}

The variable V104 was classified by Drissen and Shara (1998) as an
eclipsing binary star. Indeed the HST light curve of this star shows
two clear bumps with different amplitudes and a period of slightly more
than 12 hours, behavior suggestive of a W UMa star. Caputo et al.
(1999) also suggested that this star is an eclipsing binary and found a
period around 0.741~~d. However, from their Fig. 2 one can see that $V$
the magnitude of this star is around 15.0 with $B-V$ = 0.4.
These properties place this star inside the area occupied by RR Lyr
stars on the color magnitude diagram.
 
Our Fig. 1 shows a completely different light curve of this variable. We
phased the observations with a period of 0.31093~~d (typical of RRc
stars) but one can clearly see large systematic departures from strict
periodicity. This strongly suggests that this star is multiperiodic.
Periodograms calculated with Fourier, AoV (Schwarzenberg-Czerny 1997)
and CLEAN (Roberts et al, 1987) software are presented in Fig. 4, and 
confirm our hypothesis. There are two clear peaks in each periodogram in
this figure, one at $3.012 c/d$ (P=0.332~~d) and the second at $3.217
c/d$ (P=0.311~~d). Only one of these periods can correspond to the 1st
overtone radial pulsations. The second one is most likely connected with
non-radial pulsation -- behavior seen before only in three RRc stars in
M55 (Olech et al. 1999).
 
\section{Conclusions}
 
We have presented CCD photometry of 61 variables from the core of the
globular cluster M5 based on the newly developed method of image
subtraction (Alard and Lupton 1998, Alard 1999b). We have demonstrated
that this method works very well even in the densest regions of the
cluster. The number of variables detected in the same field using
DoPHOT photometry measured from the same data set was over two times
smaller. In addition, the relative errors of the Fourier coefficients
of the light curves measured with DoPHOT are up to 20 times larger than
the errors produced by the ISM. We conclude that the new method is very
promising for searches of variable stars in globular clusters and in
the near future should return numerous new discoveries plus very
accurate relative photometry from ground based data, even with medium
sized telescopes delivering moderate seeing. Image subtraction is best
suited for projects for which the knowledge of the zero point is not
critical, i.e.  determination of periods. On the other hand it will not
underperform DoPHOT in a sense that we may supplement accurate
difference photometry with less accurate zero points obtained with
traditional tools.
 
Four of our variables were previously unknown. One of them is a $\delta$
Scuti star with period of 0.089767~~d, one is a Bailey
type RRc star, and the other two are Bailey type RRab
stars. 
 
We have found that the variable V104, previously classified as an
eclipsing variable, is an RRc Lyr star pulsating with two periods. We
have concluded that due to the close proximity of these periods only one
of them corresponds to the 1st overtone radial pulsations and the second
one is caused by non-radial pulsations. Non-radial pulsations are
common among $\delta$ Scuti stars (the main sequence variables
laying in the main instability strip) but are rare among RR Lyr stars.
Only 3 such stars were known before, all of them RRc variables in
M55 (Olech et al. 1999).
 
\section*{Acknowledgments} 
We would like to thank to Prof. B.
Paczy\'nski for helpful hints and comments.  AO and JK were supported by
the Polish Committee of Scientific Research through grant 2P03D-011-12
and by NSF grant AST-9528096 to Bohdan Paczy\'nski.

\clearpage

\noindent {\bf Fig. 1} Light curves of the variable stars found in the
core of M5. The stars are plotted according to the increasing period. The
integral from 0 to $P$ from the light curve of each star is always zero.
 
\includegraphics{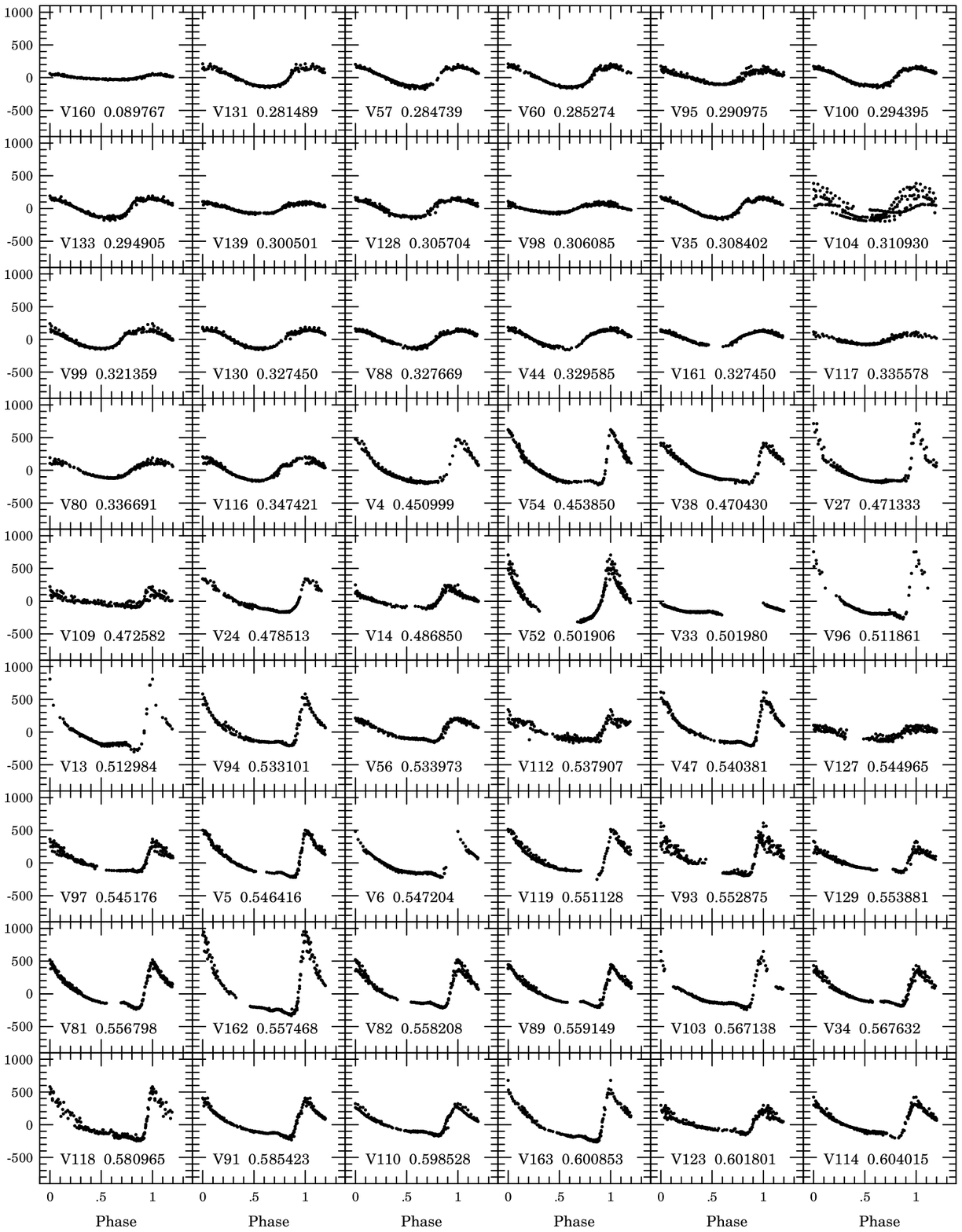}

\clearpage

\noindent {\bf Fig. 1} Continued.
 
\includegraphics{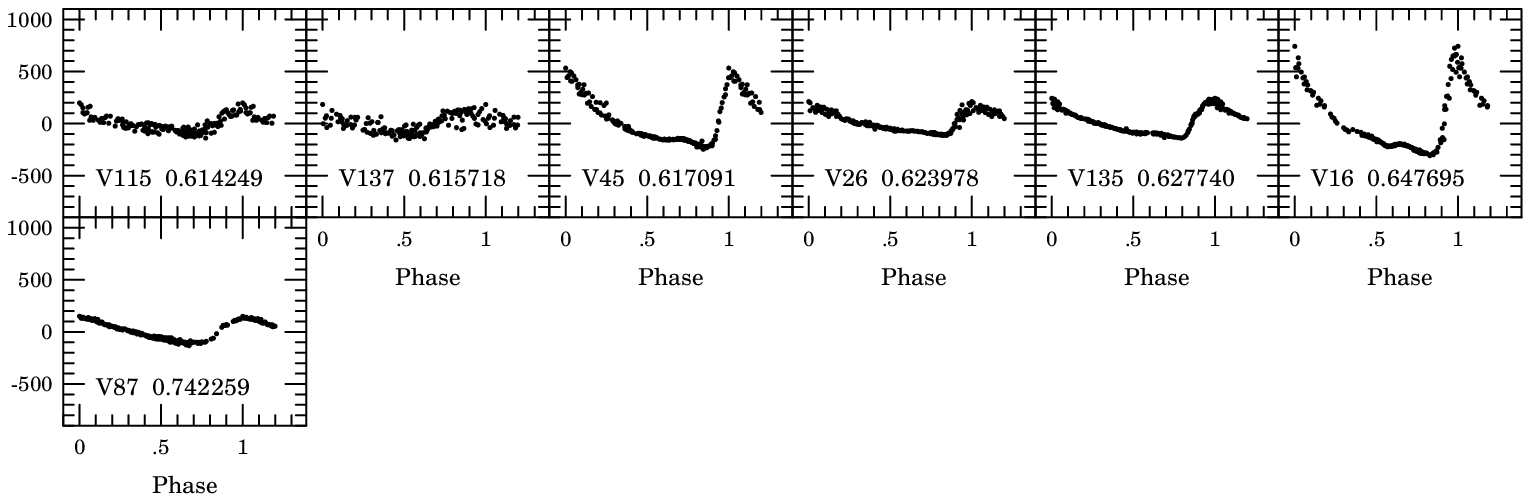}
 
\clearpage

\noindent {\bf Fig. 2} ~~Finder charts for the four newly discovered variables.
Each chart is 30 arcsec on a side, with east up and north to the left.

\includegraphics{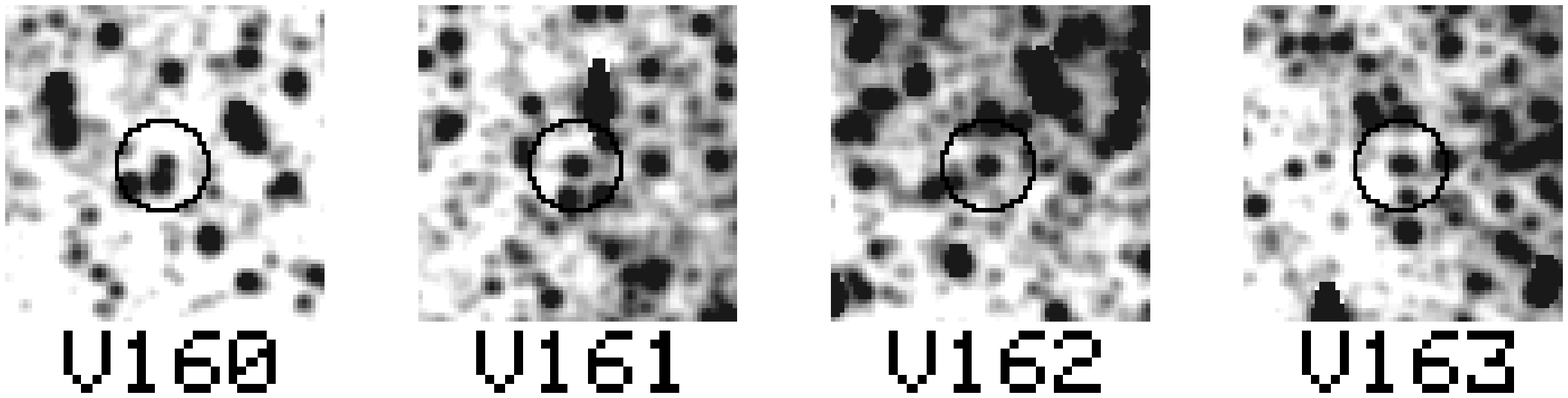}

\clearpage

\noindent {\bf Fig. 3} ~~A comparison of light curves obtained by
Kaluzny et al. 1999b using  DoPHOT software and in this work by the ISM.

\includegraphics{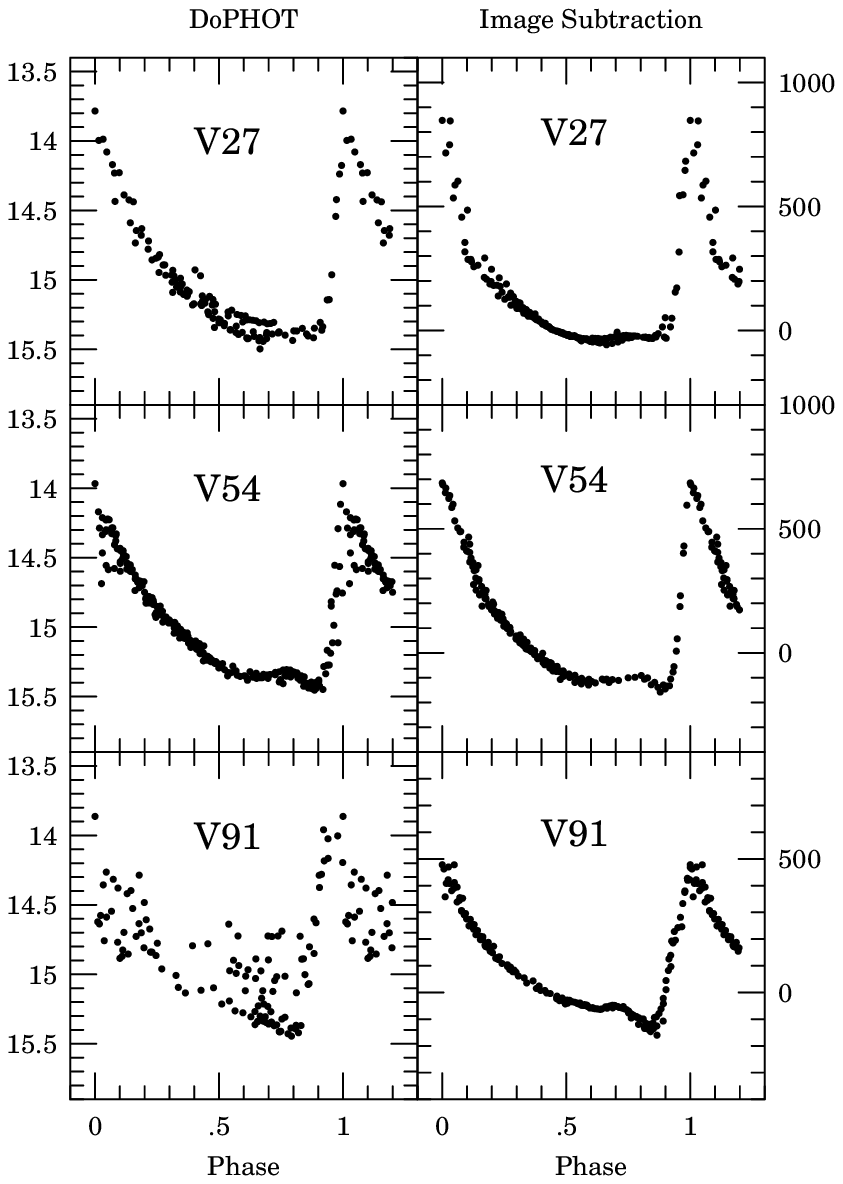}
\clearpage

\noindent {\bf Fig. 4} ~~Fourier, AoV and CLEAN  power spectra of the
light curve of V104 showing the multiperiodic behavior of this star.
 
\includegraphics{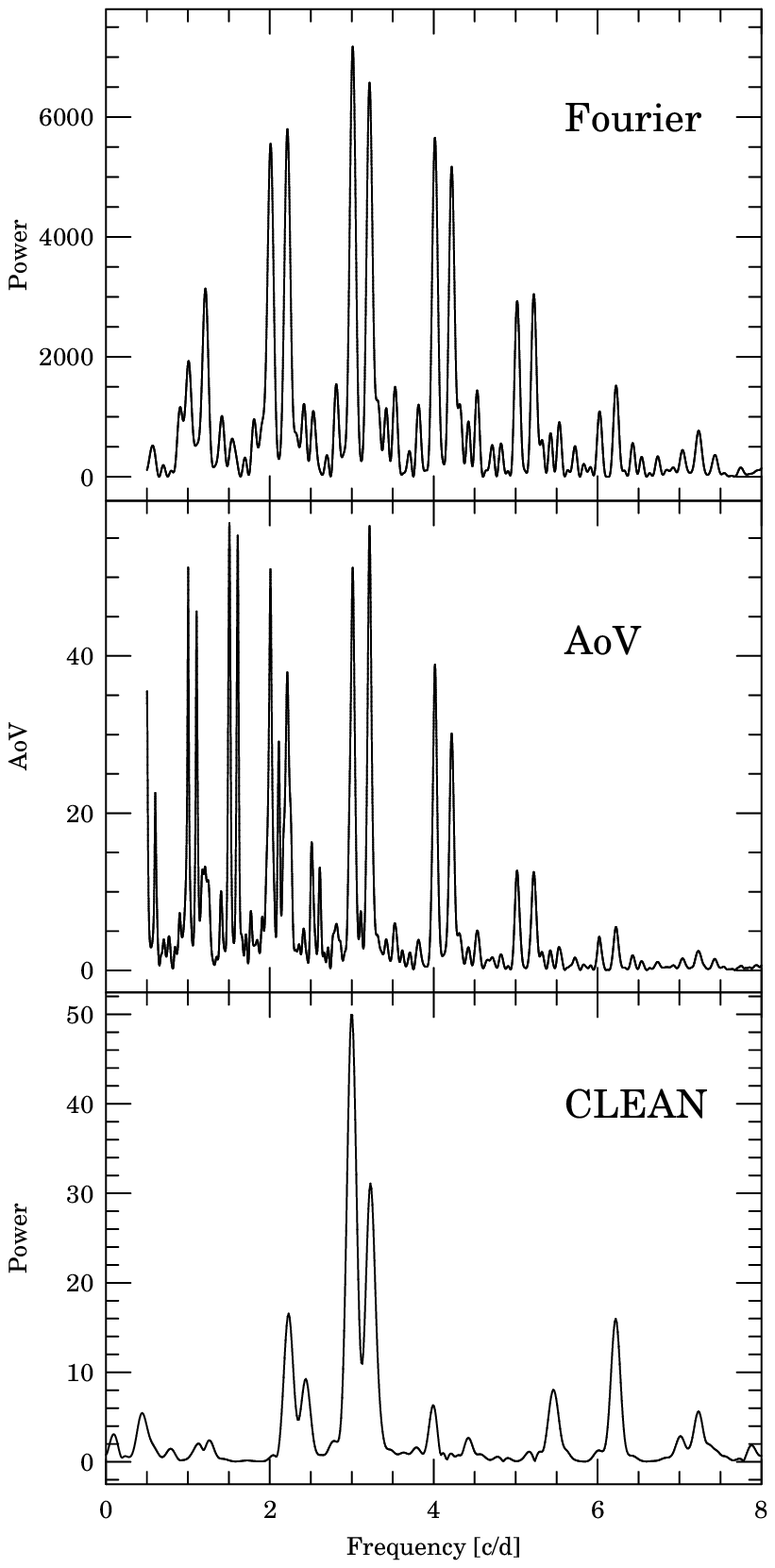}

\label{lastpage}

\end{document}